\documentclass[11pt]{article}
\usepackage{graphicx}
\usepackage{xcolor}
\def\lap{\lower.5ex\hbox{$\; \buildrel < \over \sim \;$}}
\def\gap{\lower.5ex\hbox{$\; \buildrel > \over \sim \;$}}
\usepackage{enumerate}
\usepackage{hyperref}
\usepackage{cleveref}
\usepackage{amssymb}
\newcommand\aj{Astronomical Journal}
\newcommand\araa{Annual Review of Astronomy and Astrophysics} 
\newcommand\apj{Astrophysical Journal} 
\newcommand\apjl{Astrophysical Journal Letters} 
\newcommand\apjs{Astrophysical Journal Supplements} 
\newcommand\aap{Astronomy and Astrophysics}
 
\newcommand\mnras{Monthly Notices of the Royal Astronomical Society} 
\newcommand\prl{Physical Review Letters} 
\newcommand\physrep{Physics Reports}
\newcommand\pasp{Publications of the Astronomical Society of the Pacific}
\newcommand\nat{Nature} 
 
\newcommand\prd{Physical Review D} 
 
\newcommand\aaps{Astronomy and Astrophysics Supplement}

\begin{document}
\medskip
\centerline{\bf Improving Physical Cosmology:}
\centerline{\bf An Empiricist's Assessment}
\medskip

\centerline{P.~J.~E. Peebles}
\centerline{Princeton University} 
\centerline{Princeton NJ 08544}
\centerline{June 4, 2021}
\medskip

\centerline{\it The world is so full of a number of things,}
\centerline{\it I'm sure we should all be as happy as kings.}
\centerline{\small\qquad\qquad\qquad\qquad\qquad Robert Louis Stevenson}\bigskip

\centerline{\bf Abstract}

\bigskip\noindent The $\Lambda$CDM cosmology passes demanding tests that  establish it as a good approximation to reality, but it could be improved. I present a list of possibly interesting and less well explored things that might yield hints to a better theory.

\bigskip\centerline{\bf Introduction}

\bigskip\noindent 
Our universe has ample room for a great number things, some of which are observationally accessible, and a fraction of which might teach us something new and interesting about the large-scale nature of the universe. I present a list of examples along this line in the next section. But evaluation of ideas about departures from the standard theory requires an assessment of the evidence in support of this theory. Comments about the state of the cosmological tests that establish the $\Lambda$CDM cosmology as a good approximation to reality, but certainly not a perfect one, are presented here and continue in the next section along with thoughts about how we might do even better by paying closer attention to less well-explored lines of evidence. 

I order the sources of the productive cosmological tests to date as follows: \begin{enumerate}[1.]
\setlength{\itemsep}{-4.5pt}
\item the galaxy redshift-magnitude relation;
\item stellar evolution ages;
\item abundances of the light isotopes;
\item structures, distributions, and evolution of clusters of galaxies;
\item structures and evolution of galaxies;
\item measures of the distributions of galaxies in real and redshift space;
\item measures of the CMB intensity spectrum and angular distribution.
\end{enumerate}
The last two have been particularly valuable because many of the predictions from perturbation theory are reliable; the predictions are tested by precise and well-checked measurements; and the comparisons of predictions and observations produce tight constraints on cosmological parameters that are consistent with independent measures of the cosmic mean mass density, the baryon mass fraction, the cosmological constant, the primeval abundances of helium and deuterium, Hubble's constant, and the age of the expanding universe. 

The ``Hubble tension'' is the 10\%\ difference between two independent sets of constraints on the value of Hubble's constant \cite{tensions}. One is based on observations at close to the present epoch: measurements of galaxy redshifts and distances and the relative timing of events observed in multiply lensed AGN images. The other is inferred from the theory of decoupling of baryonic matter and radiation at redshift $z\sim 10^3$. An error of 10\%\ from tracing cosmic expansion back by a factor of a thousand is impressive accuracy. There are similarly impressive degrees of consistency of constraints on the other cosmological parameters. And it is important that these constraints are based on many ways to observe phenomena set by developments at very different stages of cosmic evolution: light element formation at redshift $z\sim 10^9$, the baryon acoustic oscillation (BAO) patterns in the distributions of  the galaxies and the fossil sea of radiation set at decoupling at $z\sim 10^3$, the evolution of the cosmic expansion rate from $z\sim 1$ to $z=0$, and what is observed to be happening now. This broad variety of ways to look at the universe adds up to a compelling empirical case for the $\Lambda$CDM theory as an impressively good approximation to reality. Maybe more tensions will be found as the constraints improve. If so then I expect the case for $\Lambda$CDM as a useful approximation will remain compelling \cite{reality}, and there will be more clues to a still better theory. 

Research programs planned and in progress will improve many of the cosmological tests that brought us to the present state of cosmology \cite{2020Silk}, and we certainly hope will yield hints to how to do better. This is good science: work to improve tests that have been found to be useful. The theme of this essay is that it also is good science to look into less well explored issues that might broaden our understanding of the nature of the universe. 

\newcounter{topic}

\bigskip\centerline{\bf A Survey of Empirical Issues in Cosmology}\medskip

There is not a sharp division between research in physical cosmology and the rest of natural science; all are aspects of the universe. Consider for example simulations of galaxy formation that assume the initial conditions and dark matter physics of the standard $\Lambda$CDM cosmology, and consider also ideas about possible variants of this theory. Research groups are developing simulations that indicate the expected properties of the dark matter halos of galaxies under the assumptions of different degrees of dark matter self-interaction, or values of the dark matter particle rest mass. The results are important constraints on the nature of the hypothetical dark matter of the standard cosmology. These research groups also are examining the role of baryons in the formation of galaxies. The complexity of star formation and the effects of stars on the rest of the matter means we have no guaranty that the comparisons of simulations to observations will be a useful cosmological test, but the possibility certainly looks worth exploring. Simulations of the formation of  planets around stars also test cosmology, in principle, but I expect the interpretation is far too difficult for a meaningful cosmological test at the present state of the art. This is a matter of judgement. The same is true of the choices outlined here of adventurous and less well explored ideas for empirical research in cosmology that may prove to be interesting and productive, and perhaps even evolve into big science in a decade or two.

My assessments of issues assume the cosmological tests have convincingly established the $\Lambda$CDM theory as a useful approximation to reality: not exact, but not at all likely to be far off. This is for each to judge, of course. 

I do not comment on obviously important issues where I have no thoughts to add to  discussions in the literature. It will be fascinating to see tighter constraints on neutrino masses, for example, but I have no idea about what that might teach us about the large-scale nature of the universe. The issue of cores and cusps in the distributions of dark and baryonic matter in galaxies is very relevant to the nature of dark matter, but this issue is being thoroughly considered in theory and observations \cite{cores_cusps}. 

A final remark: depending on readers' expertise they may notice that I offer assessments of recent developments in topics in astronomy and cosmology that I do not necessarily know and understand well. I welcome advice on corrections and additions for a possible second edition.

\bigskip\noindent(\refstepcounter{topic}\thetopic\label{CosPrin}) {\it Einstein's Cosmological Principle}\medskip

\noindent The two starting assumptions for physical cosmology are that the universe is well described by Einstein's general theory of relativity and by Einstein's cosmological principle. This principle has a weak form, the assumption that the universe we can observe is pretty much the same everywhere, with no edges we can observe. The strong form is that the observable universe is well approximated as a stationary and isotropic random process with the power spectrum amplitude and tilt from scale-invariance determined by application of the $\Lambda$CDM theory to the cosmological tests.

The isotropic distributions of the X-ray and microwave backgrounds and the counts of radio sources and galaxies agree with the weak form; these were important early empirical indications that the universe is close to uniform out to observable distances. More recent deeper radio surveys continue to satisfy the weak form \cite{Flows_radio}, and the eROSITA all-sky X-ray survey \cite{eROSITA} seems to be capable of a significant addition to this test.

The strong form of the cosmological principle is checked by the cosmological tests. For example, the decoupling of baryonic matter and radiation at redshift $z\sim 10^3$ left the observed BAO statistical patterns in the spatial distribution of the galaxies and the angular distribution of the CMB. The pattern in the galaxy distribution has a characteristic length scale, the value determined by the cosmological parameters that set the boundary conditions and the speed of pressure waves in the plasma-radiation ``fluid'' prior to decoupling. This characteristic length scale predicts a characteristic angular scale in the galaxy distribution observed at a given redshift. This angular scale measured as a function of redshift at $z\lap 1$ can be compared to the prediction of the Friedman-Lema\^\i tre model with the cosmological parameters derived from the CMB anisotropy. Figure~21 in \cite{BAO_BOSS} shows the close consistency of what is measured at $z\lap 1$ and what is imprinted in the CMB at $z\sim 10^3$. This is impressive empirical support of the strong form of the cosmological principle: the distant universe, observed in different parts of the sky at $z\sim 10^3$, has to be similar enough to the relatively nearby universe observed at $z\lap 1$ to allow this close fit.

From an empirical point of view the cosmological principle says nothing about what the universe is like at distances too large to be observed, apart from indirect effects on what can be observed. To be explicit about this, consider that in standard physics there is not a locally causal relation between our peculiar velocity and the peculiar gravitational field associated with the departures from an exactly homogeneous mass distribution. Both are aspects of the growing departure from homogeneity that was set by initial conditions, presumably set at a time in the past when something like inflation allowed the causal fixing of initial conditions. This is a constraint, to be compared to one of Maxwell's equations, $\nabla\cdot\vec E=\rho$. The equations are Lorentz-covariant, though this one doesn't look it, because the value of the charge density $\rho(\vec r,t)$ at given time $t$ sets the divergence of the electric field everywhere at time $t$, not causally, but simultaneously, a constraint. An analog in the gravity physics of the relativistic cosmological model is that mass fluctuations that are outside the mass distribution that in principle can be observed after inflation ends are associated with peculiar motions that can be observed. A  mass fluctuation on a very large scale would be associated with the same motion of matter and the CMB, which is not observable, so there is no problem. But density fluctuations not so far outside our horizon can contribute to the CMB quadrupole and dipole, that is to the local motion of matter relative to the CMB. 

This situation may be relevant to the important challenge of reconciling our motion relative to the CMB with the peculiar gravitational acceleration associated with observed departures from exact homogeneity \cite{Flows_radio, supernovae, FlowsQuasars, FlowsClusters}. The measured CMB dipole anisotropy indicates the Local Group is moving relative to the CMB at about 600~km~s$^{-1}$ \cite{Kogut93}. The long history of measurements of the large-scale departures from homogeneity that would be expected to have produced this motion relative to the CMB includes a report of consistency \cite{DavisNusser}. But recent examinations  find that the gravitational attraction of observed departures from homogeneity would be expected to produce motion in about the observed direction but at considerably larger speeds \cite{Flows_radio, supernovae, FlowsQuasars, FlowsClusters}. The consistency of  results from these quite different measures of the large-scale departures from homogeneity adds to the case for the presence of an anomaly. There are challenges to be considered, however. 

First, one must find a realistic estimate of the bias in the spatial two-point position correlation function of the objects---galaxies, clusters of galaxies, quasars, radio sources, supernovae---relative to the mass  correlation function. If the sources are more strongly clustered than the mass, to a degree greater than assumed in the analysis, then the expected flow relative to the CMB is overestimated. This is in the  direction of the proposed anomaly. 

The second challenge is practical: maintain tight control of detection efficiency of objects as a function of position across a large span of the sky.  A modest systematic error could wrongly indicate a significant large-scale flow. A definitive case may have to await results of the Euclid mission \cite{Euclid} ``to capture signatures of the expansion rate of the Universe and the growth of cosmic structures: Weak gravitational Lensing and Galaxy Clustering (Baryonic Acoustic Oscillations and Redshift Space Distortion).''

Third, on the theoretical side, I have not found an analysis of the expected effect of a full spectrum of growing departures from homogeneity in producing a difference between the peculiar gravitational acceleration associated with the growing inhomogeneity and the gravitational acceleration associated with the observable part of the inhomogeneity.

If the peculiar velocity anomaly is well determined to be much larger than the predicted effect, including suprahorizon fluctuations and the effect of bias in the tracers of the mass distribution, what might be the conclusion? Since the cosmological tests make a case that persuades me that the $\Lambda$CDM theory is a good approximation to reality, I expect that the anomaly, if real, adds to clues to a still better theory. It would be an advance rather than a revolution. 

\bigskip\noindent(\refstepcounter{topic}\thetopic\label{topic:baryondensity}) {\it The Baryon Density}\medskip

We have two precision measurements of the cosmic baryon density. One is derived from the statistical patterns imposed on the distributions of the CMB temperature and polarization at decoupling at redshift $z\sim 10^3$, as predicted by the $\Lambda$CDM theory. The other is derived from measurements of the primeval abundance of deuterium and the theory of formation of the light isotopes at $z\sim 10^9$. Recent applications give
\begin{eqnarray}
&&\Omega_{\rm baryon}h^2 = 0.02237 \pm  0.00015\ \hbox{BAO  \cite{Planck_Collaboration}},\nonumber \\
&&\Omega_{\rm baryon}h^2 = 0.02166 \pm 0.00015 \pm 0.00011\ 
	\hbox{BBNS \cite{BBNS_baryon_density}}.
\end{eqnarray}
(The evidence is the Hubble parameter is close to $h=0.7$, so $\Omega_{\rm baryon}\simeq 0.045$.) The uncertainties in the Big Bang Nucleosynthesis (BBNS) measurement are written separately for the measurement of the primeval deuterium abundance relative to hydrogen and for the application of the  nuclear physics used in the conversion to the baryon density. 

The difference between the two measures amounts to about 3.3 standard deviations, which is formally significant but not interesting here because sorting out possibly quite subtle systematic errors will require more work. 

The difference amounts to three precent in the comparison of measurements based on what happened at two epochs separated by a factor of a million in the course of expansion of the universe. This is an impressive demonstration of the predictive power of the $\Lambda$CDM theory. 

The constraints on the baryon density from the situation at low redshifts are less tight, but they are important because they are independent ways to probe the universe. The estimates of $\Omega_{\rm baryon}$ from the cluster baryon mass fraction \cite{cluster mass fraction}, and from intergalactic dispersion measures from Fast Radio Burst \cite{FRBs}, agree with equation~(1) to a few tens of percent. The sum of contributions to the baryon budget from what is inferred from observations in and around galaxies amounts to 70\%\ of  equation~(1) \cite{baryonbudget}. The difference is an interesting puzzle, but the similarity of the measures in equation~(1) to this lower bound and the other two measures at low redshift is a valuable check. We can add yet another check, that the baryon density derived from absorption line spectra in the Ly-$\alpha$ forest at redshift $z\sim 3$ is comparable to the other results \cite{Meiksin}. 

The consistency of results from this broad variety of ways to observe aspects of the universe and interpret the observations by applications of different elememts of the $\Lambda$CDM theory makes an excellent case that this theory is a good approximation to what happened. If $\Lambda$CDM were wrong one might expect to encounter the occasional accidental consistency of measures of $\Omega_{\rm baryon}$, but it would be silly to imagine that the considerable string of consistent measures of the baryon density is accidental, or the result of a conspiracy. 

This argument is not a proof; we don't do proofs in natural science. We construct pictures---models or theories---and evaluate them by their ability to predict results of observations different from what went into construction of the picture. The greater the variety of successful predictions the greater the credibility of the picture. The broad variety of ways to find the baryon density, and the consistency of results within reasonable allowance for measurement uncertainties, is an impressive demonstration of the predictive power of the $\Lambda$CDM theory.
 
 A tighter constraint on the present baryon density seems feasible, if the needed resources were available, and it would be a valuable aid to tighter tests and the search for more tensions to be expected if the Hubble tension is real. 

Should we revisit the question of whether some galaxies are made of anti-baryons? The very sensitive tests for antihelium by the Alpha Magnetic Spectrometer suggests the Milky Way contains little antimatter, but could other galaxies be made of antimatter? \cite{antihelium, AMS} A sharp division of regions of matter and antimatter could invite an unacceptable domain wall, but an imaginative theorist might find a way around that. Might the primeval distributions of matter and radiation be adiabatic while the primeval ratio of the baryon number density to the dark matter density is a function of position? It would have to be carefully designed to avoid messing up the Faber-Jackson and Tully-Fisher relations, and the consistency of measures of the mean of the absolute value of the local primeval baryon number density. But there is the curiously empty Local Void to be considered in Topic~(\ref{topic:LVoid}). Might the Local Void contain the usual 10\%\ of the cosmic mass density but a smaller than usual proportion of baryonic matter?

\bigskip\noindent(\refstepcounter{topic}\thetopic\label{matterdensity}) {\it The Matter Density}\medskip

\noindent We have constraints on the matter density parameter $\Omega_m$ from six main ways to look at the universe: 
\begin{enumerate}[a.]
\setlength{\itemsep}{-4pt}
\item the old dynamical estimates \cite{Omegam_dynamics};
\item modern dynamical and weak lensing measurements \cite{DES};
\item evolution of masses of clusters of galaxies at $z\lap 0.5$ \cite{Bahcall}; 
\item supernova redshift-magnitude observations at $0\lap z\lap 2$ \cite{PerlmutterSchmidt, Tonryetal}; 
\item cross-correlation of the CMB sky temperature with the galaxy distribution, interpreted as the integrated Sachs-Wolfe effect at $z\lap 1$ \cite{ISW}; 
\item the baryon acoustic oscillation, BAO, pattern set in the CMB angular distribution at decoupling, $z\sim 10^3$ \cite{Planck_Collaboration}; 
\item the BAO pattern set in the spatial distribution of the galaxies at decoupling. \cite{BAO_BOSS}. 
\end{enumerate}
The dynamical estimate in (a) is quite approximate; I would say these measurements alone seriously constrain the mass density to $0.1\lap\Omega_m\lap0.4$. But they have been in the literature for a long time, and they continue to be a credible independent check of other ways to constrain the mass density. The Dark Energy Survey (DES) applies these old methods and adds the important constraint from weak gravitational lensing. This and other of the tighter constraints in the above list are
\begin{equation}
\Omega_m = 0.34 \pm 0.03\ (b);\ 0.28 \pm 0.05\ (d);\  0.315 \pm 0.007\ (f).
\label{eq:massdensity}
\end{equation}
The other estimates might be trusted to several tens of percent. 

The point of this discussion is the consistency of these independent ways to observe the universe from many sides and arrive at estimates of the cosmic mean mass density by application of the $\Lambda$CDM theory. If this theory were not a useful approximation to reality then there would be no reason to expect that these different approaches would yield consistent constraints. This is why constraint (a), though not at all precise, is a valuable addition to the cosmological tests of $\Lambda$CDM. 

Let us notice also that the consistency of results of these diverse probes of the universe argues that none has been seriously affected by systematic errors. That certainly may change as the tests tighten, and perhaps add to the Hubble tension. 

This evidence from measures of the matter density parameter is to be added to the case for $\Lambda$CDM from the consistency of independent estimates of the baryon density parameter $\Omega_{\rm baryon}$. At the risk of tedious repetition of the obvious, I emphasize  that these two lines of evidence together make a compelling case for the $\Lambda$CDM theory, one that good science demands be taken seriously. 

\bigskip\noindent(\stepcounter{topic}\thetopic) {\it Cosmic Helium Abundance}\medskip

\noindent Estimates of the helium abundance in the sun, planetary nebulae, and H\,II regions indicate the helium mass fraction is $Y\sim 0.3$ \cite{OsterbrockRogerson, HoyleTayler}. This local estimate of $Y$ agrees with what is inferred from the theories of decoupling at $z\sim 10^3$ and light isotope formation at $z\sim 10^9$. This result alone is impressive evidence for the $\Lambda$CDM theory. 

Given adequate resources could this test be improved by a tighter constraint on the  primeval value of $Y$ derived from astronomical observations at low redshifts? We hunger for still more demanding challenges to $\Lambda$CDM.

\bigskip\noindent(\refstepcounter{topic}\thetopic\label{timecheck}) {\it Ages of Stars and the Universe}\medskip

\noindent Since galaxies of stars are observed at redshifts well in excess of unity, the ages of the oldest stars are expected to be only slightly less than the time of expansion of the universe since it was too hot and dense for stars to have existed. I dislike being repetitive but must emphasize the impressive case for $\Lambda$CDM from the consistency of two quite different lines of evidence: stellar evolution ages  and the cosmic expansion time inferred from the theory and observations of the BAO patterns in the distribution of matter and radiation set at decoupling at $z\sim 10^3$ \cite{ages}. 

I also must repeat the formulaic question: given adequate resources could stellar evolution ages be made more accurate? It would improve an important cosmological test, and maybe reveal another tension, which would be even more important. 

\bigskip\noindent(\refstepcounter{topic}\thetopic\label{topic:DM}) {\it Dark Matter}\medskip

\noindent The consistency of the constraints on the cosmological parameters mentioned above, at the level of a few tens of percent, and based on so many different ways to look at the universe, makes a compelling case for the  $\Lambda$CDM theory and the presence of its hypothetical dark matter, DM. 

A less broadly based but maybe more direct piece of evidence for DM is the small amplitudes of the BAO patterns set in the distributions of radiation and matter at decoupling. The amplitudes are small because only the baryons, not the greater mass in DM, take part in the oscillations of the plasma-radiation fluid prior to decoupling, apart from the small pull of gravity. The unacceptably large amplitude of the oscillations in the power spectrum of the matter distribution to be expected in the absence of DM is illustrated in Figures~4 and~5 in \cite{PeebYu}.

The two immediately pressing lines of dark matter research are the search for properties of the DM and the search for its place in an improved standard model for particle physics. We may hope for hints to both from well-established programs aimed at finding signatures of DM from laboratory detectors, astronomical observations of DM annihilation or decay, manifestations of properties of DM in halos of galaxies large and small, effects of accretion of DM by the sun or earth \cite{DM in stars}, and manifestations of DM particles in high energy accelerator experiments. Something good will come of all this.

What do we make of the similar cosmic mass densities in baryons and DM? We could live on a planet in a solar system in a universe with the observed mass density but all baryons, no dark matter. We could equally well live in a universe with the standard mass density and a much smaller baryon mass fraction. Maybe the similarity of the mass densities is only an accident, though it seems unlikely. Or maybe a more complete particle theory will require this near coincidence.

\bigskip\noindent(\stepcounter{topic}\thetopic) {\it Einstein's Cosmological Constant}\medskip

\noindent We have evidence of detection of Einstein's Cosmological Constant $\Lambda$ from four different lines of observation:
\begin{enumerate}[a.]
\setlength{\itemsep}{-4pt}
\item the BAO signature in the galaxy spatial distribution;
\item the BAO signature in the CMB angular distribution;
\item the supernova redshift-magnitude relation; 
\item the comparison of stellar evolution ages and the cosmic expansion time.
\end{enumerate}
I mentioned in Topic~(\ref{CosPrin}) the proposal that large-scale cosmic flows could cause a significant systematic error in the measurement of the SNe1a redshift-magnitude relation \cite{supernovae}. If so it would vitiate this line of evidence. But as indicated in Topic~(\ref{matterdensity}) the evidence for the presence of $\Lambda$ from the patterns left from decoupling on the distributions of matter and radiation, items (a) and (b) in this list, is tested by the theory and observation of a considerable number of bits of evidence. These measurements alone make a serious case for the presence of something that acts like $\Lambda$. 

Items (a) and (b) are independent tests based on quite different methods of probing the universe, the first using observations of the spatial distributions of the luminous galaxies in samples at redshifts $z\lap 1$, the second measurements of the angular distribution of the CMB radiation that propagated to us almost undisturbed since redshift $z\sim 10^3$. The case for the presence of $\Lambda$ from the CMB  measurements was already clear by the year 2000, but this required that Hubble's constant is reasonably close to what is found from local observations of galaxy redshifts and distances. Later observations of the effect of weak gravitational lensing on the pattern of the CMB anisotropy enabled the establishment of the presence of $\Lambda$ from the CMB measurements alone \cite{CMBweaklensing}. 

We have yet another independent piece of evidence for the presence of $\Lambda$ from the consistency of the cosmic expansion time with stellar evolution ages (Topic~\ref{timecheck}). If $\Lambda$ were not taken into account it would make the ages of old stars greater than the age of the expanding universe. 

These tests assume Einstein's general theory of relativity and the cosmological principle, with the postulates of dark matter and the cosmological constant, and some assume Hubble's constant is reasonably close to the astronomers' measurement. But if these assumptions were seriously wrong, or the $\Lambda$CDM theory were wrong, the consistency of these quite demanding observationally independent tests would be truly remarkable, far more unlikely than most of us would be willing to consider. The sensible conclusion is instead that we have a compelling case for the presence of $\Lambda$ in the $\Lambda$CDM theory.

The immediately pressing challenges are on the theoretical side: find the places for $\Lambda$ and the quantum zero-point energy density in an improved standard physics. So we have issues.

\begin{enumerate}[a.]
\item The quantum zero-point energy of matter, and the interaction of this zero-point energy with gravity, are real; both pass serious experimental tests. For example, to get the computed binding energy of molecular hydrogen to agree with the measured value requires taking account of the zero-point energy associated with the degree of freedom of the distance between the two hydrogen atoms when they are in the bound ground level. And the experimentally tested equivalence of energy and active gravitational mass therefore means the zero-point energy of matter is a source of gravitational acceleration. Is the zero-point energy density of the electromagnetic field also real and a source of gravity? If not it would call for an agonizing reappraisal of what we have found to be reliable physics.

For another way to put the situation consider the S-matrix operator connecting initial and final states of a system in quantum field theory. The perturbative expansion of the S-matrix represented as a sum of Feynman diagrams includes closed loops with no external legs. They contribute to the vacuum energy density. To represent the active gravitational mass of a material object we have to imagine adding to these vacuum loops external legs to represent the active gravitational mass of an object that is a source term in Einstein's field equation. But if material fields have these gravity legs then they make a ridiculously large contribution to the vacuum stress-energy tensor that is the active gravitational source in Einstein's field equation. One can add gravity legs to loops for boson fields, which helps reduce the sum, but not nearly enough. 

If the vacuum zero-point energy density of the quantum fields is real, and the numerical value is independent of the motion of the observer, then it acts as $\Lambda$. This is elegant, except that the value would be expected to be absurdly large. If the field zero-point energy is real and violates Lorentz symmetry the reconciliation with particle physics is another problem to be added to the absurd energy density \cite{Lorentz_symmetry}. 

\item Must we resort to the anthropic principle \cite{Weinberg89} to account for the value of $\Lambda$ in the standard cosmology? It has been pointed out that if there is a multiverse of universes, each with its own variety of physics, then the anthropic choice of a universe similar to ours, with a similarly small absolute value of the vacuum energy density, is inevitable. This discussion continues in Topic~\ref{AnthropicPrinciple}. 

The search for an explanation of the quantum vacuum, by the anthropic principle or by a better approximation to the physics of our own universe continues, as in~\cite{Arkani_Hamedetal}. Further  close analysis of this challenging issue will teach us something of value.

\item The search for evolution of $\Lambda$ \cite{RatraPeebles} is good science. So is the search for evolution of the strength of the electromagnetic interaction, $e^2/\hbar c$, notably by John Webb and colleagues \cite{fine_structure_constant}. There is a well-supported Dark Energy Survey, DES \cite{DES}. Why isn't there a well-supported Fine-Structure Survey, FSS? 

This line of thought began with the proposal that the strength of the gravitational interaction, $Gm^2/\hbar c\sim 10^{-38}$, where $m$ is the nucleon mass, is so small because it has been slowly evolving to zero for a long time \cite{Dirac, Dicke1961}. The Lunar Laser Ranging experiment has established that the strength of the gravitational interaction has not been evolving faster than one percent of the rate of expansion of the universe \cite{LLRanging}. Who knows what this might portend for DES and FSS?

\item In the standard cosmology we flourish not long after the cosmological constant and the mass density in matter made equal contributions to the expansion rate. This curious, one might say unlikely, coincidence used to be considered a good argument against $\Lambda$ and for the scale-invariant Einstein-de~Sitter model. That argument has been empirically falsified. I have not encountered  an anthropic explanation of the coincidence. Must this be dismissed as just another unlikely coincidence?

\end{enumerate}

\bigskip\noindent(\refstepcounter{topic}\thetopic\label{topic:GravPhysics}){\it Gravity Physics}\medskip

\noindent\label{GravityMOND} Einstein's general theory of relativity, GR, passes precision tests on scales ranging from the laboratory to the solar system, length scales $\lap 10^{13}$~cm. The theory passes the considerable array of cosmological tests of its application on the scale of the Hubble length, $\sim 10^{28}$~cm. It certainly is important to explore the consequences of modifications of GR, but the thought that GR does so well on such an impressive range of scales, and starts to require modification just when reaching the scales of cosmology, seems a little contrived. And since the standard model for the dark sector---nonbaryonic dark matter and $\Lambda$---is a lot less thoroughly tested my impression is that research into the properties of the dark sector is a more likely near-term bet for improvement.

Would a different gravity theory and a cosmology without dark matter, built along the lines of Milgrom's MOND \cite{topicMOND}, pass the considerable array of tests of gravity physics and cosmology? It seems exceedingly unlikely. Consider for example the difficulty of accounting for the small amplitudes of the BAO patterns mentioned in Topic~\ref{topic:DM} in the absence of dark matter.  I conclude that we have a convincing argument for the presence of something that acts like dark matter. More comments on this judgement are in Topic~\ref{topic:Galaxies}\ref{onMOND} on the properties of large galaxies. 

\bigskip\noindent(\stepcounter{topic}\thetopic) {\it The Spatial Distribution of the Galaxies}\medskip

\noindent The cosmological tests that establish the $\Lambda$CDM theory depend on reliable measurements, but the reliability of the predictions is equally essential. We have many reliable observations of galaxies, but their use as cosmological tests is limited by the complexity of star formation and the effects of stars on their environments. The character of the space distribution of the galaxies is an intermediate situation; the computation of predictions requires approximations but  their reliability does not seem to be seriously disturbed by complexity. 

An important example from 1997 is the prediction by Neta Bahcall {\it et al.} that in the Einstein-de~Sitter model the masses of rich clusters of galaxies would grow more rapidly than observed \cite{Bahcall}. This was early credible evidence that the mass density is less than in the Einstein-de~Sitter model that was popular then, and the evidence remains credible. The combined applications of theory, numerical simulations, and observations of clusters of galaxies has grown into a rich subject \cite{clusters2021}, and will grow even more rich with the Euclid project \cite{Euclid}. 

Other aspects of the galaxy spatial distribution are considered in the next several topics.

\begin{figure}[h]
\begin{center}
\includegraphics[angle=0,width=3.in]{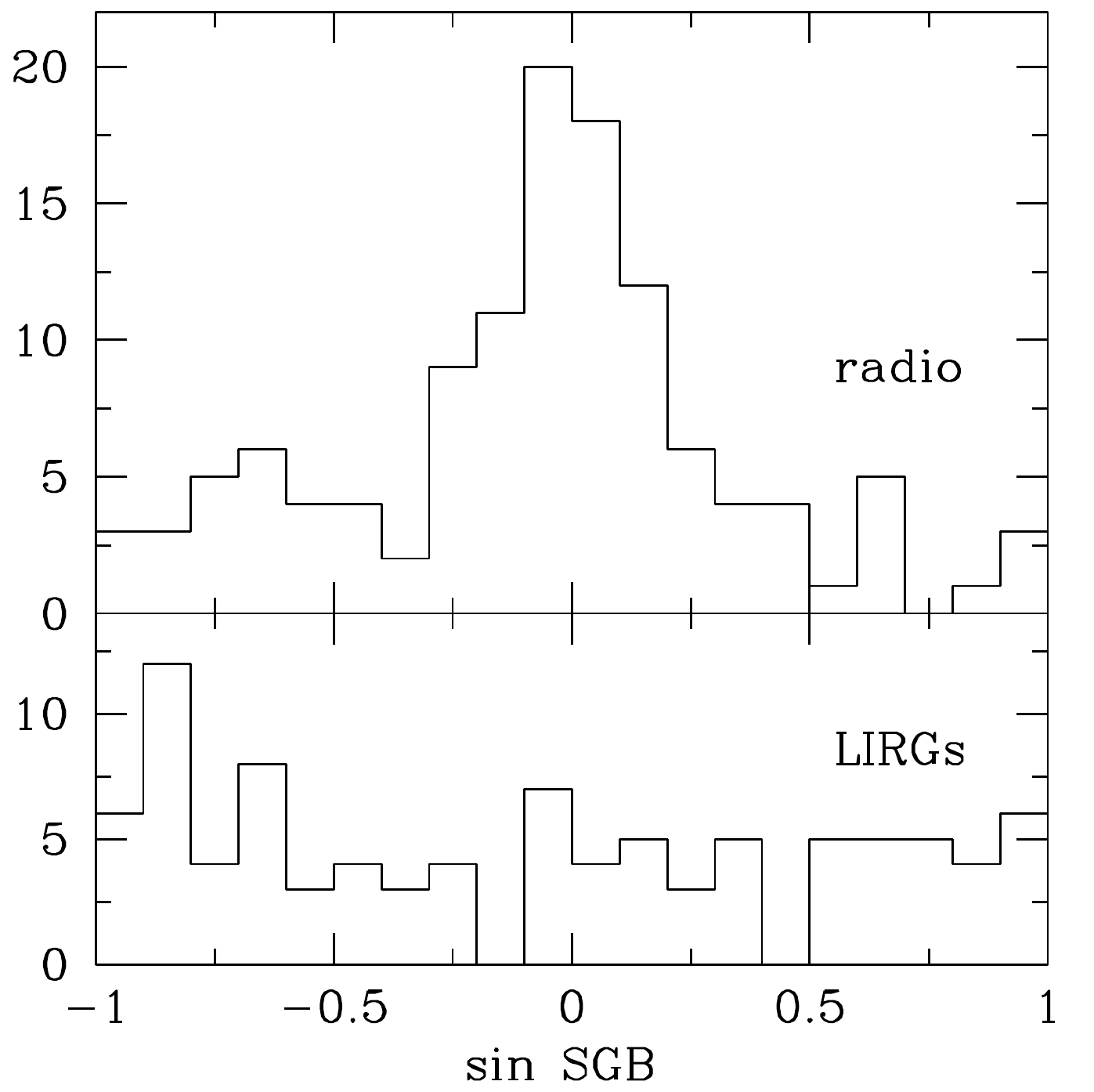}
\caption{\small Distributions of angular positions of two kinds of galaxies, those unusually luminous in radio, and those unusually luminous in the infrared. Both are at redshifts $z< 0.02$. In the horizontal axis SGB is the latitude relative to the plane of the Local Supercluster defined by the galaxies at $D\lap 10$~Mpc. An isotropic distribution has on average equal counts in equal intervals of $\sin({\rm SGB})$. Galaxies that are luminous at radio wavelengths prefer the plane; galaxies that are luminous in the infrared do not.\label{Fig:histograms}}
\end{center}
\end{figure}

\bigskip\noindent(\stepcounter{topic}\thetopic) {\it The Extended Local Supercluster}\medskip

\noindent An aspect of the galaxy distribution that might become an interesting cosmological test, and just possibly big science,  builds on the observation that most of the nearby galaxies, at distances $D\lap 10$~Mpc, are close to a plane that G\'erard de Vaucouleurs named the Local Supercluster. In the late 1980s Brent Tully \cite{Tully(1987)} noticed that distant clusters of galaxies tend to be concentrated to the direction of this plane of the Local Supercluster, and Peter Shaver \cite{ShaverPierre, Shaver} independently noticed this effect for clusters and for luminous radio galaxies within redshift $z = 0.02$, distance $D=85$~Mpc. Pierre and Shaver pointed out that at this distance the large galaxies show no readily perceptible preference for the plane of the Local Supercluster. A later reexamination shows that optically selected galaxies do show a marginally significant excess density at low supergalactic latitudes extending out to about 50~Mpc \cite{ORS_LSC}.

Figure~\ref{Fig:histograms} shows an updated illustration of the effect. The two histograms show counts of objects at redshift $z<0.02$ in equal intervals of $\sin({\rm SGB})$, where SGB is the supergalactic latitude. These are equal solid angles of the sky, so an isotropic distribution would show similar counts in each bin, with the usual fluctuations due to the correlated, clumpy, distributions of the objects. 

The bottom histogram shows the angular distribution of the 93 galaxies detected in the Infrared Astronomical Satellite (IRAS) sky images at 12, 25, 60 and $100\mu$, and entered in the PSCz catalog of detections that have point-like infrared images and optical spectra identifiable as galaxies \cite{PSCz, LIRGs}.\footnote{I use the list at \url{https://heasarc.gsfc.nasa.gov/W3Browse/all/iraspscz.html}, with the choice of parameters: source sample main, redshift status 1, optical class code g.} These galaxies are exceptionally luminous in the infrared, and known as LIRGs. Their distribution in $\sin({\rm SGB})$ has considerable fluctuations, an effect of the clustering tendency of galaxies, but there is no indication of a preference for directions in the extended plane of the Local Supercluster. 

The top histogram shows the distribution of the 121 radio-luminous galaxies at redshifts $z<0.02$ in the all-sky catalog compiled by Sjoert van Velzen et al. \cite{radiogals}. These radio galaxies are distinctly concentrated toward the extended plane of the Local Supercluster, as Shaver and Pierre saw in 1989. The 16 clusters of galaxies at $z<0.02$ listed by Struble and Rood \cite{StrubleRood} also are concentrated toward this plane, as Tully noticed. The radio galaxies in clusters of galaxies are a subset of this sample; we see that many more radio galaxies are in groups that also prefer the plane.

I have not taken into account the interference to the discovery of LIRGs by infrared emission by dust in the plane of the Milky Way, and the interference to the discovery of radio galaxies by obscuration of optical images by dust and confusion of radio detection by local radio sources, all to be required in a more formal analysis. But the interference does not look to be a serious problem because the plane of the Milky Way is tilted almost $90^\circ$ from the plane of the Local Supercluster, so the confusion along the zone of avoidance is spread across most of the bins in SGB. 

The evidence from the figure looks clear, as it was three decades ago. At distances $D<85$~Mpc galaxies that are exceptionally luminous at radio wavelengths tend to be close to the plane of the Local Supercluster that is so well defined by the galaxies at $D<10$~Mpc, along with the Virgo cluster at $D=20$~kpc. But LIRGs, which are exceptionally luminous at wavelengths $\sim 60~\mu$, show no preference for the plane. The infrared radiation from LIRGs is produced by interstellar dust that is thought to be heated either by rapid star formation or by AGNs. If the latter it would be a curious difference from the distribution of AGNs identified as radio galaxies.  

Little has been made of the striking contrast between the distributions of galaxies that are exceptionally luminous in the infrared and those exceptionally luminous at radio wavelengths. This is not surprising because it is difficult to know what to make of it. That is no excuse, though; the effect seems quite real so we are challenged to investigate it. My choice for the first order of business is to check for  other possible features of this extended plane of the Local Supercluster. Do early-type galaxies at $D < 85$~Mpc that are not in clusters prefer this plane? Might spirals avoid the plane? What is the situation at $D > 85$~Mpc? A larger project, maybe a meet use of AI, is the search for other preferred planes in the spatial distribution of what is now a large sample of clusters of galaxies with their radio sources.

\bigskip\noindent(\refstepcounter{topic}\thetopic\label{topic:LVoid}) {\it The Local Void}\medskip

\begin{figure}[ht]
\begin{center}
\includegraphics[angle=0,width=4.5in]{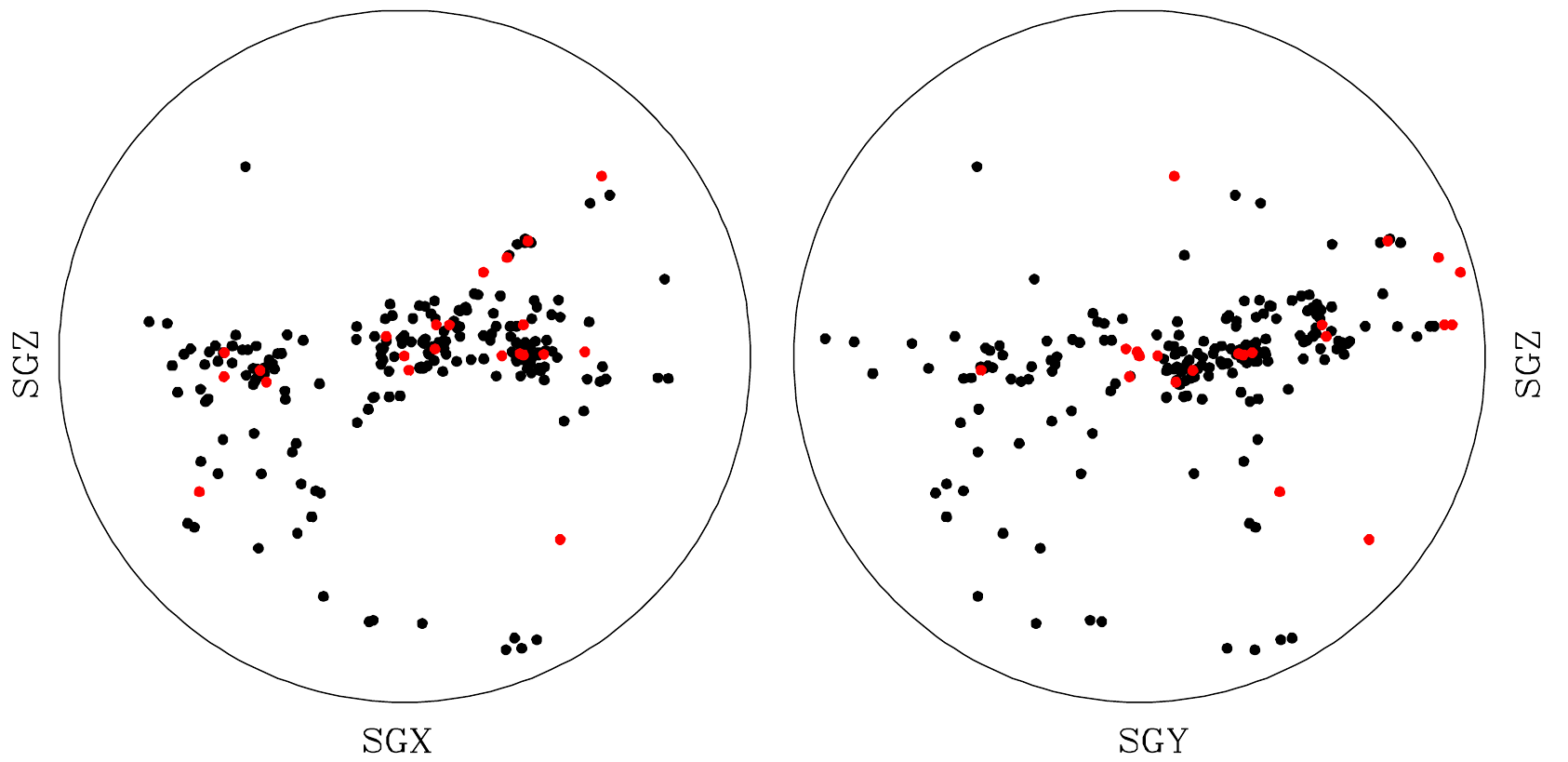}
\caption{\small The distribution of galaxies within 8~megaparsecs, in orthogonal projections in supergalactic coordinates.\label{fig:LUmap}}
\end{center}
\end{figure}

\noindent Just one dwarf galaxy, KK\,246 = ESO 461-36, at distance $\sim 7$~Mpc, is known to be well inside the Local Void that occupies one third of the volume out to 8~Mpc distance \cite{PeeblesNusser, KK246Courtois}.  The situation is illustrated in  Figure~\ref{fig:LUmap}.\footnote{\label{fn:LUcatalog}The data for this figure are available at the Tully et al. Extragalactic Distance Database, {http://edd.ifa.hawaii.edu} as the catalog ``Local Universe (LU)''. I made this figure in 2013; I don't think it has been previously published. An earlier version is in \cite{PeeblesNusser}.} The red symbols are the 22 large galaxies, $L_K>10^{10}$, and the black symbols the 200 tabulated less luminous galaxies at distances $1 < D < 8$~Mpc and luminosities $10^6 < L_K < 10^{10}$. That is, these maps exclude the dwarfs near the Local Group but include the Milky Way and M31. 

The figure shows the extreme range of  local densities of galaxies on the scale of several megaparsecs: the concentration of galaxies in the plane of the Local Supercluster and the much lower densities above and below the plane. The remarkable fraction of this 8~Mpc sample in the almost empty region above the plane, the near part of the Local Void, is best appreciated by comparing the orthogonal projections.

Numerical pure dark matter simulations indicate that the DM density in voids in the galaxy distribution is roughly 10\%\ of the cosmic mean. Tinker and Conroy \cite{TinkerConroy} showed that the use of the Halo Occupation Distribution prescription to translate halo masses in  simulations to galaxy luminosities gives a reasonable account of these close to empty nearby regions. But is it sensible to think that the formation of stars and HI clouds in this Local Void is so efficiently suppressed? Could the Local Void really be so empty? \cite{voidphenomenon}

The image of KK\,246 is easy to see on the digitized ESO sky survey  photographic plates (when I have been shown where to look). The HI cloud around the stars in KK\,246 was detected by the Parkes HIPASS survey that scanned the sky at just 450~seconds per beam element \cite{HIPASS}. Surveys planned and in progress with technology capable of detection of far fainter optical and HI sources will give a lot more information about what is and is not in the Local Void. But the question is interesting enough to merit a Grand Project to supplement the broad surveys. It would be fascinating to see the results of the deepest practical  21-cm examination of the sky in a redshift range centered on KK\,246, 430~km~s$^{-1}$, and covering a modest part of the sky centered on KK\,246. I leave the choice of ranges of angular positions and redshifts to be sampled in this project to more capable hands. 

\bigskip\noindent(\stepcounter{topic}\thetopic\label{NActionM}){\it Tracing Galaxy Positions Back in Time}\medskip

\noindent The Numerical Action Method (NAM) finds solutions to the equation of motion of galaxies back in time under mixed boundary conditions: growing peculiar velocities at high redshift, as befits the primeval departures from homogeneity, and the best approximation to observed present distances, redshifts, and proper motions where available. NAM does a reasonably good job of finding cosmologically acceptable solutions \cite{NAM}. It would be interesting to repeat the analysis for the many more dwarfs out to 5~kpc or maybe further that could be detected and for which there could be useful measures of distances and redshifts.  
 
The motions of the Magellanic Clouds and the Magellanic Stream traced back in time are well studied \cite{Magellanic}, but I have not seen checks of peculiar velocities of model orbits at redshifts greater than unity. This is important because the uncertainties of position and velocity computed from the present conditions grow larger as the path is computed back in time. This means that the path computed back in time is likely to have the galaxy moving at quite unacceptably large  speed at high redshift. The NAM analysis that takes this into account is presented in \cite{NAM_Magellanic}. It certainly could be done better.

\bigskip\noindent(\refstepcounter{topic}\thetopic\label{topic:Galaxies}){\it The Properties of Large Galaxies}\medskip

\noindent Some observations: Estimates of the cosmic mean mass density in stars and the star formation rate density as functions of redshift indicate that  ``galaxies formed the bulk (75\%) of their stellar mass at $z < 2$'' \cite{stellarmassdensity}. At a given present stellar mass, spirals formed their stars later than ellipticals. A more familiar way to put this is that spirals are still forming stars while ellipticals are red and dead. Since the ratio of ellipticals to spirals is smaller at lower mass, another way to put it is the Cowie et al. downsizing effect \cite{downsizing96, downsizing}: lower mass galaxies on average formed the bulk of their stars later. Here are some issues.

\begin{enumerate}[a.]

\item The Milky Way galaxy is particularly interesting because it can be examined in greatest detail, with much more to come from Gaia \cite{Gaia}. 

I am instructed that the MW stellar halo is not a mixture of stars similar to those observed in MW satellites, unless they were assembled at a very early stage in the formation of the stars \cite{Tolstoy}. The observations are consistent with the formation of the MW stellar halo in one piece that would have been salted later by stars from accreted dwarfs that produced the ``field of streams'' \cite{fieldofstreams}. 

\item\label{onMOND} The MOND \cite{topicMOND} proposal to replace the dark matter of the standard model with a modified form of the gravitational central force law predicts the baryonic Tully-Fisher relation between the circular velocity in a galaxy and the mass in stars plus gas. The prediction is a remarkably good fit to the observations (Fig.~4 in \cite{McGaugh}). This result must have something of value to teach us. What might it be?

Although the dark matter postulated in the $\Lambda$CDM cosmology allows us to understand the rotation curves of spiral galaxies and the relative motions of the galaxies, these observations might be equally well explained by a MOND-like cosmology. The situation with respect to the cosmological tests is quite different, however. The results reviewed above, with the abundance of well cross-checked measurements, makes a compelling empirical case for $\Lambda$CDM with its dark matter. Could a MOND-like cosmology without dark matter do as well? The idea that a seriously different theory would pass the broad range of cosmological tests seems exceedingly unlikely. That is a judgement of plausibility, of course, not a  proof, but good science demands that the abundance of evidence that fits the $\Lambda$CDM theory be taken seriously. 

\item Pure disk galaxies. Some nearby spiral galaxies, such as M\,31 and M\,81, have classical bulges of stars centered on the galaxy and rising out of the disk in the manner of an elliptical galaxy. But the majority of the $L\sim L_\ast$ galaxies closer than 10~Mpc are spirals, and most of the spirals have inconspicuous classical bulges \cite{Kormendy2010}. You can see their images on the web, most in beautiful detail. I particularly recommend the  image of M\,101, which shows its spiral arms running into a nuclear star cluster, you might term it a classical bulge, with luminosity no more than $10^{-4}$ times the luminosity of the galaxy. The MW is in this class; it has a peanut-like bar but little starlight in a central bulge. 

These pure disk galaxies had to have grown by accretion of diffuse baryons---gas or plasma---with minimal accretion of stars that would add to a bulge/halo. If they formed by mergers of subhalos, as seen in numerical simulations of galaxy formation, then star formation had to have been almost entirely confined to one of the subhalos. How could this have been arranged? Why the one favored subhalo?  
  
\begin{figure}[h]
\begin{center}
\includegraphics[angle=0,width=4.5in]{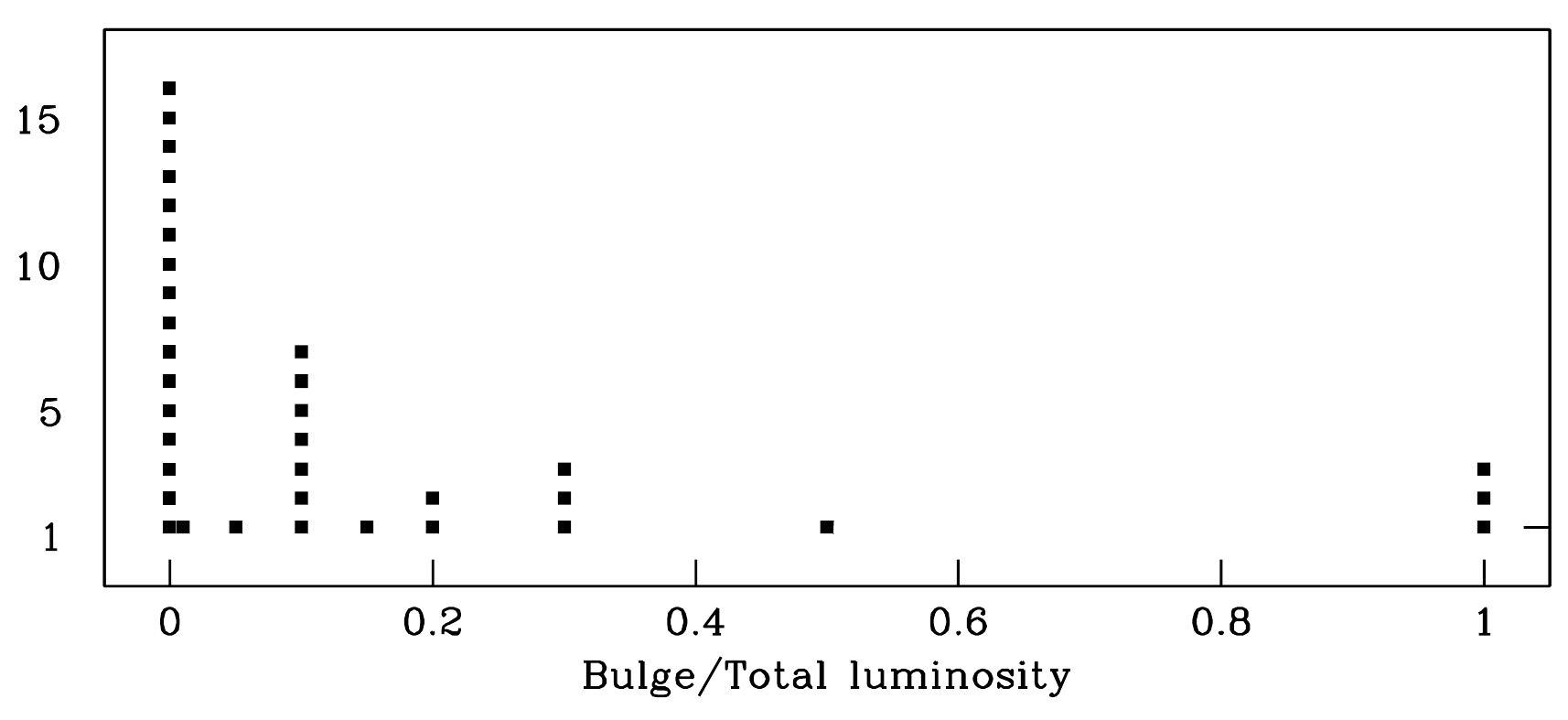}
\caption{\small Frequency distributions of ratios of bulge to total luminosities of nearby $L\sim L_\ast$ galaxies.\label{BtoT} }
\end{center}
\end{figure}

\item\label{bimodality} Galaxy bimodality is seen in scatter plots of galaxy color and optical luminosity: most galaxies are in the red sequence or the blue cloud \cite{color_magnitude}. Bimodality is seen also in different star formation histories in spirals and ellipticals, and related to that the different relative abundances of iron and the $\alpha$-elements. The phenomenon is observed most directly in the distinctly bimodal character of the images of the nearby $L \sim L_\ast$ galaxies you can see on the web. 

Most of the nearby large galaxies are of two quite distinct types: spirals or ellipticals, disk-dominated or bulge-dominated, star-forming or quenched, late or early types. Figure~\ref{BtoT} shows the frequency distribution of the ratios $B/T$ of bulge to total luminosities  of the 32 galaxies within 10 Mpc distance and luminosities $L_K > 10^{10}$ for which there are measured values of $B/T$ in \cite{Kormendy2010} or \cite{FisherDrory2011}. Most are spirals, and at least half of them have at most inconspicuous bulges. There are just three ellipticals: Centaurus\,A, Maffei\,1 and M\,105. They look distinctly different from most of the other galaxies, though there are exceptions.  NGC\,2784 is an S0 that looks like an elliptical to my untutored eye, but might be the remnant of a spiral that was somehow stripped of its gas and dust. The Sombrero Galaxy looks like a large elliptical that contains a large spiral. NGC\,4490 looks like it is merging or falling apart. And the NASA/IPAC Extragalactic Database seems uncertain about the classification of the Circinus Galaxy. But these exceptions are not common in the nearby sample.  If the three dozen or so nearest $L\sim L_\ast$ galaxies are a fair sample of the situation outside clusters, as suggested by the bimodality seen in color-magnitude plots for much larger samples of galaxies, then the evidence is reasonably clear; large galaxies exhibit a distinct bimodality. How did this come about?

It is said that elliptical galaxies formed by dry mergers, spirals by wet. Perhaps evidence of this is seen in the Centaurus group with its two large galaxies. The satellites around the elliptical Centaurus A are largely early types, and the satellites around M\,83, a spiral with an inconspicuous bulge, are almost all late types \cite{Centaurus}. Thus we can put the question two ways. Why is there a prominent bimodality of galaxy morphologies? Or why did some galaxies form by mergers of dry subhalos, others wet, in what looks like a bistable process? 

In a familiar bistable situation a separatrix marks the boundary of parameter values that separate evolution of the system in one of two directions, to one mode or the other. For example, it is generally thought that the giant $L\sim 10L_\ast$ cD galaxies in rich clusters formed by mergers of many cluster members. In this picture we may say that environment is the separatrix between the formation of $L\sim L_\ast$ galaxies and formation of the giant $L\sim 10L_\ast$ cDs. 

Among the two most common $L\sim L_\ast$ types an obvious difference is that spirals are supported largely by rotation and ellipticals are not. (S0s are a complication, but they're not common nearby.) This invites the speculation that the transfer of angular momentum to a protogalaxy was bistable, a large or small amount. However, there is no evidence of this in pure DM simulations. The ratio of numbers of the two dominant types, early and late, is an increasing function of the stellar mass of the galaxy, in simulations and observations \cite{early_late}. But there are nearby large and small ellipticals and large and small spirals, so the protogalaxy mass alone cannot define the separatrix. Ellipticals prefer denser regions, but there are $L\sim L_\ast$ ellipticals well away from clusters, in our neighbourhood. We might suppose, maybe in desperation, that all protogalaxies began to evolve toward the spiral morphology, but that some violent disk instability turned some protodisks into protoellipticals. 

Here is a reasonably well established phenomenon, galaxy bimodality. It presents us with a fascinating opportunity for research: identify the natures of the bistability and the separatrix. 

\item\label{mergerremnants} Merger remnants. The disks of present-day spirals are subject to a violent instability:  merging of $L\sim L_\ast$ galaxies. The web shows clear examples; the Antennae Galaxies is a stunning illustration. What do these merger remnants become as they relax to a close to steady state? The issue has a long history  \cite{JPO_on_mergers}. Remnants of serious mergers certainly are not pure disk spirals; there would be far too many bulge/halo stars. I think not normal ellipticals; wrong star formation history, wrong iron to $\alpha$-element abundance ratio. Maybe something similar to the Sombrero Nebula or the field S0 galaxy NGC 3115? Are there  $L\sim L_\ast$ ellipticals close enough to be examined in the detail needed to to establish that they are remnants of spirals that merged after they had become recognizable spirals?

There is ample evidence of more distant ellipticals and spirals that have tails and streams of stars that look very much like results of mergers \cite{vanDokkum, stellarstream}. I have not found estimates of the fraction of galaxies that exhibit these symptoms, however.  If the local galaxies are close to a fair sample of the situation outside clusters, then relaxed $L\sim L_\ast$ merger remnants are not common, because pure disks are common. I have not found how this compares to the relative abundance of remnants of mergers of recognizable spirals in simulations.

\item Elliptical galaxies as island universes.\label{E_islands} The fundamental plane in the  space of elliptical galaxy luminosity, radius, and velocity dispersion is strikingly insensitive to the abundance of neighboring galaxies \cite{BernardiFP}. The mean spectra of ellipticals are systematically different at different stellar velocity dispersions, but again at given velocity dispersion the mean spectrum is almost the same for ellipticals in more crowded and less crowded environments (apart from more  prominent H-$\alpha$ emission in field ellipticals) \cite{EspectraZou}. If ellipticals grew by dry mergers of star clusters that had a range of values of velocity dispersions, and hence a range of different spectra, then one might have predicted a sensitivity of the assembled elliptical to the present environment, which might be expected to have correlated with the degree of merging. Mergers certainly happen, and tidal tails are clear evidence of it. But the effect is hard to see in the spectra. 

In the scatter plot of galaxy color and absolute magnitude the ratio of counts of red to blue galaxies increases with increasing galaxy luminosity, but again the position of the red sequence in the color-magnitude plot is quite insensitive to environment \cite{redsequence}. This is not what one might have expected if red galaxies grew by mergers of less luminous early type galaxies that would have had less red colors if they had evolved in isolation. Here again the degree of assembly by mergers had to have been almost indifferent to environment.

There are environmental effects. Bernardi {\it et al.} (2006) found that the surface brightnesses of ``galaxies in dense regions tend to be 0.08 mag fainter than those in the least dense regions'' \cite{BernardiFP}. This is an effect of environment, but impressively modest. McDermid {\it et al.} (2006) \cite{McDermid_downsizing} found that galaxies with larger stellar mass formed their stars earlier, an example of downsizing. They also found that, at given stellar mass, galaxies in the relatively dense Virgo cluster formed stars earlier than ellipticals of similar mass in the rest of their sample, which are in lower density regions. But we see again the interesting situation. Figure~18 in \cite{McDermid_downsizing} shows that the correlation of age with stellar mass is pronounced, and the correlation with environment relatively modest. 

So ellipticals are not strictly islands; their properties are correlated with environment; but the correlations are small compared to the variation with luminosity, mass, or velocity dispersion. From an empirical point of view, the hierarchical growth of an $L\sim L_\ast$ elliptical, as in a merger tree, is not suggested by the observations reviewed here. That is not to say that the merger tree concept is misleading, but that it is a possibility to be kept in mind. 

\item Spiral galaxies as island universes. I have not seen similar checks of spirals but can note two things. First, the spirals M\,101 and NGC\,6946 are close to the Local Void seen in Figure~\ref{fig:LUmap}, yet they look like the normal grand design spirals seen in the crowded plane of the Local Supercluster. And second, the pure disk spirals that are so common nearby cannot have gained significant luminosity by accretion of satellites containing stars, because that would have added stars to their stellar bulge/halos, which are inconspicuous in most nearby large spirals. Pure disk galaxies do accrete companions, as must be expected from classical mechanics. A classic example is the pure disk edge-on galaxy NGC\,5907 with its stellar stream that we may expect eventually will add to the stellar halo of this galaxy \cite{N5907}. It happens, but certainly not often among nearby pure disk spirals. As for ellipticals, the observations do not suggest hierarchical growth of $L\sim L_\ast$ pure disk galaxies. 

\item \label{luminosity_segn} The relative distributions of large and small galaxies. Supergiant cD galaxies prefer the dense environments of clusters of galaxies. The standard and reasonable thought is that the dense environment encourages growth of these extra-large early-type galaxies by merging and ram pressure stripping, as discussed in item~\ref{bimodality}. 

\begin{figure}[h]
\begin{center}
\includegraphics[angle=0,width=5.in]{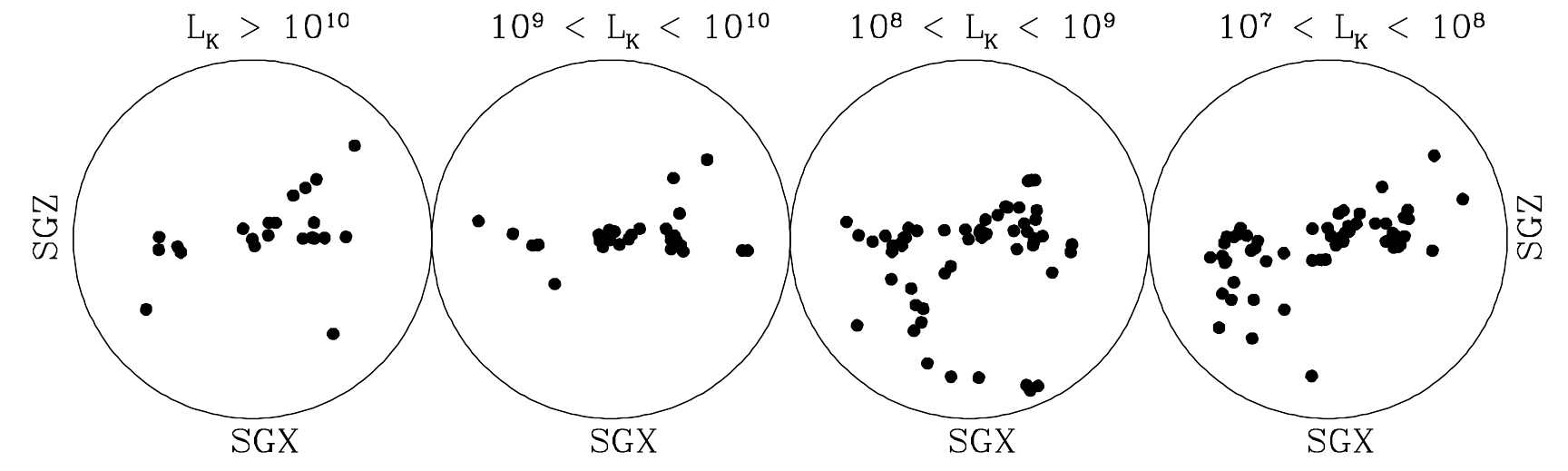}
\caption{\small Distributions of the galaxies closer than 8~Mpc in separate ranges of K-band luminosity, plotted in supergalactic coordinates.\label{fig:liminosities}}
\end{center}
\end{figure}

In the less dense range of environments outside clusters the distributions of large and small galaxies are fairly similar. This is illustrated in Figure~\ref{fig:liminosities}. The data are the same as in Figure~\ref{fig:LUmap} (from the Local Universe catalog referred to in footnote~\ref{fn:LUcatalog}). Here the dwarfs in the Local Group are plotted, but the lower bound on the K-band luminosity eliminates the extreme dwarfs that can be found only nearby. A first impression is that there are  more dwarfs than giants below the plane, as one would expect, but there are more giants than dwarfs above the plane. But considering the statistical noise in this relatively small sample, the figure does not suggest that the ratio of large to small galaxies is particularly sensitive to the local mean galaxy density. The same point was made long ago by the similarity of the maps of positions of large and small galaxies in the CfA redshift survey, Figures 2a and 2d in \cite{CfA_Maps}. 

A statistical measure of how the galaxy spatial distribution depends on luminosity is the galaxy two-point position correlation function $\xi_L(r)$ computed as a function of the galaxy luminosity $L$. Let the measurements in each bin of $L$ be fitted to the power form $\xi_L(r) = A_L\, r^{-\gamma}$ at separation $r < 10$~Mpc. For simplicity use the same value of $\gamma$ for all luminosity bins, and find the best value of the clustering amplitude $A_L$ for each bin in luminosity $L$. Results along this line \cite{Norberg2002, Zehavi2011} show that, at luminosities greater than the Milky Way, $A_L$ increases rapidly with increasing $L$, at the most luminous galaxies reaching the clustering length of clusters of galaxies. This is expected because, as was noted, $L\sim 10 L_\ast$ galaxies prefer to be in clusters of galaxies with their large clustering length. But at the less luminous end, at $L\lap L_\ast$, the clustering amplitude $A_L$ is close to independent of the intrinsic luminosity. This is in line with the impression I get from Figure~\ref{fig:liminosities}.

A standard argument leads to the expectation that at formation the local ratio of the numbers of small and large galaxies is sensitive to the local mean density. To recall the argument suppose that, for simplicity, the primeval departures from homogeneity may be pictured as the sum of a large-scale variation of density with position added to a component that varies with position on the smaller scale characteristic of a galaxy. If a positive small-scale peak happens to be present near a maximum of the large-scale component then the peak might be expected to grow into a large galaxy. The same small-scale peak that happens to be present near a minimum in the large-scale component might be expected to grow into a dwarf. That is, this simple picture invites the prediction that at formation the ratio of dwarfs to $L_\ast$ galaxies increases with decreasing local mass density, or decreasing galaxy number density. 

The present galaxy distribution would have been shaped by the gravitational growth of the clustering of the galaxies, which would have drawn together giants and dwarfs alike.  That could account for the mix of galaxy luminosities in the plane of the Local Supercluster. But if gravity grew the galaxies together to form this plane then the debris left above and below the plane would have been from the regions of particularly low local primeval density. Below the plane there are more dwarfs than giants, as expected, but above the plane we see more giants than dwarfs, an odd situation.   

The argument for luminosity segregation is sensible but seems wrong, according to the evidence reviewed here. What might this mean? It is to be added to the list of tensions, this one to be explored and perhaps resolved by simulations.

\item Do all galaxies have dark matter halos? Dwarfs formed from tidal streams may not \cite{tidal_dwarf}, but there are other more puzzling examples. NGC\,3115 is a large S0 galaxy in the field. If it has a DM halo with mass typical of its stellar mass then the halo must be much less dense than usual, the halo core radius much larger \cite{N3115}. The satellite NGC1052-DF4 \cite{vanDokkum2019} of the elliptical NGC\,1052 has seven of its own satellites that look like globular star clusters. One is at 7~kpc projected distance from NGC1052-DF4, four at 3 to 5~kpc, and the other two are at closer projected distances. The dispersion of their redshifts relative to NGC1052-DF4 is no more than 6~km~s$^{-1}$. The mass of NGC1052-DF4, based on the spread of redshifts and distances of these globular clusters, is about what would be expected if NGC1052-DF4 contained no dark matter. This does not look at all likely to be a tidal dwarf situation. The small motions of the globular clusters at considerable distances from the stars in NGC1052-DF4 instead argue for minimal tidal influence on a system that has little dark matter within the distances of the globular clusters from this dwarf galaxy. 

\item Workhorses. Simulations of galaxy formation gave us the concept of galaxy assembly by a hierarchical merger tree. This picture agrees with the fractal nature of the galaxy distribution on scales $\lap 20$~Mpc, which one might imagine reflects the hierarchical gravitational growth of clustering of the galaxies. The merger tree for the growth of galaxies has been termed one of the workhorses of cosmic structure studies, along with halo occupation distributions and abundance matching.

It is in the spirit of empiricism to ask whether, absent simulations but given our knowledge of the phenomenology, people would have been led  to employ these workhorses. On the face of it, the Eggen, Lynden-Bell, and Sandage \cite{ELS} picture of the formation of the Milky Way by near monolithic collapse fits the observations reasonably well. In this contrafactual history we would not have the guidance of merger trees, and instead a more straightforward account of the formation of the stellar halo of the Milky Way \cite{Tolstoy}. The ELS picture also  affords a simpler account of the formation of pure disk galaxies, because it avoids the problem with the stars that could have formed in subhalos before merging, stars that are unwanted because they would end up in the galaxy stellar bulge/halo. We would not have the power of abundance matching, but this operation does have the flavor of circularity. When I look at maps of observed galaxy positions the halo occupation distribution concept does not come to my mind. And I have mentioned the question of the merger tree concept for elliptical galaxies in topic~\ref{topic:Galaxies}\ref{E_islands}. 

The lesson I draw is that it is good science to employ these workhorses, and it is good science also to question them on occasion.

\end{enumerate}

\bigskip\noindent(\refstepcounter{topic}\thetopic\label{topic:simulations}){\it Numerical Simulations of Galaxy Formation}\medskip

\noindent Simulations of galaxy formation under the assumption of the $\Lambda$CDM physics and initial conditions have made impressive progress in accounting for galaxy structures, luminosity functions, star formation histories, and spatial distributions as functions of redshift. The use of $\Lambda$CDM is natural and sensible, and the encouraging results from the simulations add to the case for this cosmology. But it is sensible also to bear in mind that the theory is incomplete, so it is important that ideas about the properties of dark matter and initial conditions are being tested by their computed effects on galaxy formation. Here are some other issues to consider.

\begin{enumerate}[a.]
\item Pure disk galaxies. Simulations of the formation of large galaxies with masses comparable to the Milky Way usually produce spirals with prominent classical bulges of stars that rise out of the disk in the center of the galaxy. The nearby large spirals M\,31 and M\,81 have significant bulges, with bulge to total luminosity ratios $B/T\sim 0.3$, which is in line with the simulations. But as discussed in Topic~(\ref{topic:Galaxies}) this is contrary to the images of many of the other large galaxies that are close enough to be observed in some detail. By my count, among measured ratios of bulge to total luminosities \cite{Kormendy2010, FisherDrory2011} of the galaxies at distance $D<10$~Mpc, 11 of the 20 with luminosity $L_K>10^{10.5}$ look like pure disks with bulges that are too inconspicuous to be observed. Of the 19 with luminosities $10^{10.0}<L_K<10^{10.5}$, five  have inconspicuous classical bulges, and two more are assigned $B/T=0.01$ and 0.05, which are pretty inconspicuous.

The galaxy simulation community has not manifested much concern about this pure disk phenomenon. This is in part because the community is continuing to improve the simulations, which may continue to bring them closer to the observations. But the classical bulge issue has been in the literature for a decade \cite{Kormendy2010}, and simulated galaxy images continue to have prominent bulges. 

Another consideration is that the pure disk phenomenon is readily observed only among the closest galaxies. It is appropriate to pay attention to statistical measures from observations of enough galaxies to reach a fair sample, but one must bear in mind that this yields reliable statistical measures from observations that may be overly schematic. It is a matter of judgement to strike the practical balance between observations of enough galaxies to approach a fair sample but few enough that each can be observed in enough detail to assess issues at hand, here the phenomena of classical bulges and pure disks. But however that balance is found it surely makes sense to pay attention to the evidence from the several dozen nearest large galaxies to be seen and admired on the web, the galaxies in our face.

\item A commonly used measure of the velocity dispersion of star particles moving in the plane of a model disk galaxy is the circularity parameter $\epsilon$. We need not bother with the definition, and note only the situation in a spiral galaxy with a flat rotation curve. A model star particle moving in the plane of this disk with circularity parameter $\epsilon = 0.8$ has a distinctly eccentric orbit, with the ratio of largest to smallest galactocentric distances $r_{\rm max}/r_{\rm min}=2.6$ \cite{PeebPureDisks}. Typically at least half the orbits in a simulated galaxy are  more eccentric than this. I do not know whether this situation could be consistent with the beautiful fine structure of the spirals in M\,101, but it seems unlikely. I do not know why the galaxy formation community is not manifesting greater concern about the issue of hot model disks.

\item The galaxy bimodality phenomenon is discussed in Topic~\ref{topic:Galaxies}\ref{bimodality}. Numerical simulations produce a satisfactory fit to the observed bimodal distribution of galaxy colors as a function of stellar mass \cite{model_bimodality}. I have not seen evidence that simulations produce the pronounced distinction between disk galaxies with inconspicuous bulges and elliptical galaxies with inconspicuous disks.  A first step would be a demonstration that $L\sim L_\ast$ model galaxies have a bimodal distribution of bulge-to-total luminosities, along the lines of Figure~\ref{BtoT}.

\item Do model dwarf galaxies prefer regions of low mean density, while model $L\sim L_\ast$ galaxies avoid them? I mentioned in Topic~\ref{topic:Galaxies}\ref{luminosity_segn} that pure thought predicts it but that the empirical evidence I have seen do not support it. 

\end{enumerate}
\bigskip\noindent(\stepcounter{topic}\thetopic){\it Isolated Dwarf Galaxies} \medskip

\noindent A dwarf further than about $1$~Mpc from the nearest large galaxy seems likely to have evolved in isolation since the primeval mass distribution broke apart into gravitationally bound protogalaxies. This is by analogy to the Milky Way and M\,31; they are at present separation $0.75$~Mpc and they are approaching each other for the first time. Isolated dwarfs that have not been near a large galaxy since the protogalaxies separated might be expected to have interesting histories to compare to what happened to dwarfs orbiting large galaxies. Here are some considerations.

\begin{enumerate}[a.]

\item The dwarf galaxy KK\,246 seems to have evolved in splendid isolation in its lonely position in the Local Void (Topic~\ref{topic:LVoid}). So why is the distribution of stars in KK\,246 tilted from the H\,I cloud that surrounds it \cite{KK246}? A first thought is that two dwarfs merged to form KK\,246. To explore this it would be helpful to know the streaming motions of the stars in this galaxy to compare to the redshift pattern in the HI. Are the rotation axes of stars and H\,I not the same? Is this dwarf close enough, at $D\sim 7$~Mpc, for a measurement of the red giant branch and maybe more of the HR diagram for hints to the star formation history? Is the H\,I envelope of this galaxy likely to be settling and feeding star formation? Surely that would have tended to erase the tilt of the stars relative to the H\,I?

\item Other nearby isolated dwarfs have their stories to tell. The rule among larger dwarfs is that isolated ones are star-forming \cite{Geha}, while less isolated dwarfs tend to have been quenched by interaction with massive galaxies. Consistent with this, dwarfs near enough to the Milky Way for measurements of the stellar color-magnitude diagram show signs of bursts of star formation that might be ascribed to the disturbance by close passage of  the Milky Way or some other galaxy. But consider three nearby but pretty isolated dwarfs. 
\begin{enumerate}[i.]
\item The nearest large galaxy to the dwarf KKR\,25 is M\,31, 1.4~Mpc from KKR\,25; the closest known dwarf is  KK\,230 at 1.0~Mpc away \cite{KKR25}.
\item The nearest large galaxy to KKH\,98 is IC\,10, 1.8~Mpc away; the closest known dwarfs are IC\,5152, GR\,8, and KKR\,3, at $\sim 1.0$~Mpc \cite{MelbourneKKH98}. 
\item I think the nearest large galaxy to KKH\,18 is M\,81, 1.3~Mpc away \cite{isolated_dwarfs, KKH18}. 
\end{enumerate}
The images of these three objects look to me like dwarf spheroidals, with no obvious signs of dust revealed by obscuration or scattered starlight. All three have detected HI halos, suggesting isolated evolution appropriate for their isolated present positions \cite{KKH18, HuchtmeierKK}, though I don't know whether the HI is contributing to star formation. The image of KKH\,18 does show bright blobs of something. KKH\,98 may have had a starburst about $5\times 10^8$ years ago \cite{MelbourneKKH98}. Maybe, like KK\,246, KKH\,98 is the remnant of the merger of two dwarfs that left a dwarf spheroidal with an HI halo. Here is a fascinating subject for a serious campaign of observations and theory in near-field cosmology \cite{near-field_cosmology, near-field_cosmology2}.

\item I mentioned in Topic~\ref{NActionM} that NAM solutions for the paths of presently isolated dwarfs computed back in time to cosmological initial conditions at the separation of matter into protogalaxies yield reasonably good fits to measured redshifts and distances \cite{NAM}. Surveys in progress and planned seem sure to find many more of these nearby isolated dwarfs. It will be a rich data set for exploration of the varieties of star formation histories and for a NAM analysis of the motions of these test particles back to potentially interesting measures of initial conditions.
\end{enumerate}

\bigskip\noindent(\stepcounter{topic}\thetopic){\it Massive Black Holes}\medskip

\noindent Larger galaxies contain central compact objects with masses in the range $10^6$ to $10^{10} M_\odot$. The one in the center of the massive elliptical M~87 certainly is compact \cite{ETH_M87}; it looks like an excellent bet that this one and by extension the others are black holes. What does this mean? There are thoughts  \cite{merging_massive_BHs, merging_massive_BHs_2}.

\begin{enumerate}[a.]
\item We seem to need two relations between the central black hole mass and the stellar bulge/halo, because some of the  pure disk galaxies that are so common nearby have massive central black holes with at most modest classical bulges. The familiar example is the Milky Way, with its central bar, not much starlight in a bulge, and a central compact object with the mass of $4\times 10^6 M_\odot$. Another is NGC\,4945~\cite{N4945}. It is seen near edge-on, and looks wonderfully flat, with little starlight rising out of the disk in a stellar bulge/halo. The evidence is that this galaxy contains an active galactic nucleus presumably operating around a massive black hole with mass comparable to the object in the center of our Milky Way galaxy.  

\item How did these massive black holes in the centers of galaxies grow? Maybe they are primeval, results of some sort of rare cataclysmic events during the long span of logarithmic time in the early universe. 

The other main line of thought is that the massive central black hole grew as the galaxy grew. Perhaps the central black hole grew by the dissipative settling of a considerable mass of diffuse baryons that contracted to a black hole that continued to grow by accreting ever more matter. That growth would be expected to have released a lot of energy, which could serve to suppress the rate of star formation that in simulations without some sort of feedback effect tends to be unrealistically rapid. 

Since it is difficult to imagine that diffuse matter would settle to such a high density without fragmenting into stars, another thought is that the massive central black holes grew by the merging of smaller seed black holes. These seeds may be primeval, or may have formed in the first generation of gravitationally bound clouds of baryons with Jeans masses on the order of $10^6 M_\odot$. Maybe the central matter in some clouds settled to form $\sim 10^5 M_\odot$ seed black holes \cite{first_massive_BHs}. The objection to this is again the extreme contraction, from a cloud density of maybe $\sim 10^{-25}$ g~cm$^{-3}$ to the density characteristic of a $10^5 M_\odot$ black hole, $\sim c^6G^{-3}M^{-2}\sim 10^8$ g~cm$^{-3}$. The contraction by eleven orders of magnitude in radius would offer ample room for the growth of primeval departures of the cloud from spherical symmetry, to be added to the effect of interaction of a contracting mass concentration with all the other mass concentrations in the cloud. It seems reasonable to expect the inevitable result is multiple fragmentation and the formation of a star cluster \cite{fragmentation}. 

These different scenarios suggest different patterns of formation of gravitational waves during relativistic merging events, which if detected would be a great help in sorting out ideas about a fascinating phenomenon. Where the central massive concentrations of galaxies in the centers of galaxies come from?

\item Large spirals merge; we see the local examples discussed in Topic~\ref{topic:Galaxies}\ref{mergerremnants}. A relaxed merger might be a spiral with a prominent stellar bulge of the stars present before the merger, or might be an elliptical with an unusual pattern of chemical element abundances. Since galaxies with multiple nuclei are not common we must expect that the two massive black holes in a merger remnant typically would find a way to merge, producing a gravitational wave that is considered detectible. 

\end{enumerate}

\bigskip\noindent(\stepcounter{topic}\thetopic){\it The circumgalactic/extragalactic medium}\medskip

\noindent  Apart from historical reasons, is baryonic extragalactic matter to be distinguished from circumgalactic matter? The hint I take from the observations is that the circumgalactic matter around a large galaxy sprawls out in an increasingly irregular way with increasing distance from the galaxy until it becomes parts of the circumgalactic matter around nearby large galaxies. The result would be the clumpy extragalactic medium. An argument along this line goes as follows.

The galaxy-galaxy two-point correlation function \cite{xi_gg}, and the galaxy-mass cross-correlation function derived from weak lensing observations \cite{xigrho}, are well approximated by quite similar power laws, 
\begin{equation}
\xi_{gg}= \left(r_{gg}\over r\right)^{1.77}\hspace{-6mm},\ \ \ \ r_{gg} = 6.7\hbox{ Mpc}, \quad
\xi_{g\rho}= \left(r_{g\rho}\over r\right)^{1.8}\hspace{-5mm},\ \  \ \ r_{g\rho} = 7.3\hbox{ Mpc},
\end{equation}
for $H_o=70$~km~s$^{-1}$~Mpc$^{-1}$ and $\Omega_m=0.30$, and at separations $r$ from a few tens of kiloparsecs to about 10~Mpc. 

At separations $\lap 10$~kpc the galaxy-mass function $\xi_{g\rho}$ is the typical run of mass density $\rho(r)$ as a function of radius $r$ in a galaxy. In the power law model $\xi_{g\rho}\propto r^{-\gamma}$ this mass distribution translates to the rotation curve 
\begin{equation}
v(r)^2 ={3\over 2(3-\gamma)}\Omega_m H_{\rm o}^2r_{g\rho}^\gamma r^{2-\gamma},
\quad v(r) = 280\, \left(r\over 3\hbox{ kpc}\right)^{0.1}\hbox{ km s}^{-1}.
\label{eq:vc}
\end{equation}
The slow variation of $v(r)$ with radius $r$ is typical of the outer parts of galaxies. The galaxy-galaxy relative velocity dispersion increases with increasing projected separation $r_p$ in the range $10\hbox{ kpc} \lap r_p \lap 1\hbox{ Mpc}$ (in Fig.~6 in \cite{DavisPeeb1983}) about as expected from the scaling with radius in equation~(\ref{eq:vc}).

If the power law scaling of $\xi_{g\rho}(r)$ with radius $r$ continues to radii less than the optical sizes of galaxies then the value of the circular speed $v(r)$ at $r\sim 3$~kpc in equation~(\ref{eq:vc}) represents the mix of galaxy masses used in the measurement of $\xi_{g\rho}(r)$. This numerical value is not unreasonable.

A related measure is the ratio $v_{\rm opt}/v_{200}$ of the circular velocity $v_{\rm opt}$ within a spiral galaxy to the circular velocity $v_{200}$ of a test particle circling the galaxy at radius $\sim 200$~kpc. The analysis \cite{vg/v200} uses $v_{\rm opt}$ derived from the galaxy luminosity using the Tully-Fisher relation, and  $v_{200}$ derived from weak lensing of Sloan Digital Sky Survey galaxies. The study~\cite{vg/v200} measured the ratio for spiral galaxies with stellar masses centered on 0.6, 2.7 and $6.5\times 10^{10}M_\odot$. In all three samples 
$v_{200}/v_{\rm opt}\sim 0.7$. The scaling in equation~(\ref{eq:vc}) suggests $v_{200}/v_{\rm opt}\sim 1.5$. But the weak lensing measure would have been applied to spirals without close neighbors, while the measured galaxy-galaxy relative velocity dispersion includes galaxies with close neighbors, which increases the dispersion. That is, the two measures of $v_{200}/v_{\rm opt}$ are not manifestly inconsistent.

The similarity of the correlation functions $\xi_{gg}(r)$ and $\xi_{g\rho}(r)$ would be expected if the baryonic mass of the universe were concentrated in the halos of galaxies in the same manner as the mean concentrations of galaxies around galaxies.\footnote{This statement would be accurate if the mass of each galaxy were concentrated in the luminous part. It is a reasonable approximation on scales of a few megaparsecs, where an $L\sim L_\ast$ galaxy typically has many neighbors.} It invites the speculation that the cross-correlation function of galaxies and the circumgalactic matter is similar to $\xi_{g\rho}(r)$ and $\xi_{gg}(r)$, because the gas and plasma around a galaxy spread out to become the clumpy distribution of the circumgalactic/intergalactic matter around neighboring galaxies, in the fashion of the galaxy distribution. 

The picture finds some support from the observations of Mg\,II, Ly\,$\alpha$, and other absorption lines in the spectra of background quasars and galaxies. The redshifts of the absorption lines produced by the foreground matter correlate well with galaxies at similar redshifts and closer to the line of sight to a background quasar than roughly 100~kpc. And the redshifts often are correlated with more than one galaxy, as you would expect of a clustering hierarchy \cite{CircumgalacticMedium, MenardZhu}. 

Another way to put this picture is that most baryons are bound to the gravitational potentials of the mass in and around the concentrations of galaxies (with exceptions: cosmic rays, maybe galactic winds). Here are some other considerations.

\begin{enumerate}[a.]

\item  Does the circumgalactic/extragalactic medium extend into the Local Void?  As discussed in Topic~(\ref{topic:LVoid}) simulations put the mean density of matter, including baryons, in voids at about 10\%\ of the cosmic mean, and the issue is whether the Local Void really contain this much baryonic matter with so little of it observed. If this matter contained only the light elements produced at high redshift then the spectra of background quasars on lines of sight passing through the Local Void would show only the Ly-$\alpha$ line produced by neutral matter. KK\,246 in the Local Void has an H\,I halo that would produce a Ly-$\alpha$ line if there were a conveniently placed background source. Maybe other H\,I clouds  in the Local Void could be observed above the atmosphere, provided obscuration by the local Ly-$\alpha$ resonance absorption is not too strong at about the redshift of KK\,246.

\item The surface density in circumstellar clouds of cool gas at distances $\sim 100$~kpc from a large galaxy has similar average values in early and late type field galaxies. At smaller distances, $\lap 50$~kpc, there is considerably more cool gas around late types, as in emission line galaxies. Perhaps this is related to the feeding of, or the winds from, the greater star formation rate among late type galaxies \cite{MenardZhu, circumgalactic2011, AnandNelsonKauffmann2021}. Might it be a hint to the bistability discussed in Topic~\ref{topic:Galaxies}\ref{bimodality}?

\item Simulations suggest the circumgalactic medium near galaxies is still accreting cool streams free of chemical elements heavier than those produced in the early universe. Consistent with this, Lyman-$\alpha$ absorbers with very little evidence of heavy elements are observed at $z\gap 3$ \cite{primevalHI}. Is there evidence of this phenomenon at low redshift? Always important to test simulations.

\item Are the heavy elements in the Ly$\alpha$ forest salted by winds from star-forming galaxies or produced by local supernovae? How does either affect the spatial distribution of the HI forest relative to the dark matter mass distribution?
\end{enumerate}

\bigskip\noindent(\stepcounter{topic}\thetopic){\it Evolution of the circumgalactic/extragalactic medium}\medskip

\noindent This is probed in many ways: the Gunn-Peterson effect, absorption lines in the spectra of background quasars and galaxies, redshifted 21-cm emission and absorption, the optical depth $\tau$ for scattering of the CMB by free electrons,  maybe FRB dispersion measures as a function of redshift (Topic~\ref{topic:baryondensity}), with more to come from the Euclid and CMB-S4 experiments \cite{Euclid, CMB-S4}. These measurements are being thoroughly discussed; a few points might be noted here. 

\begin{enumerate}[a.] 
\item The optical depth for Thompson scattering of the CMB, integrated through the time it would have suppressed the CMB anisotropy, is put at $\tau\sim 0.05$ \cite{Planck_Collaboration}. This amounts to the surface mass density $\Sigma=\bar m\tau/\sigma_{\rm T}\sim 0.1$~g~cm$^{-2}$. I cannot think of an anthropic explanation of the oddly human scale of the surface density $\Sigma$. 

\item The residual ionization in the ``dark ages,'' before reionization, must  have contributed little to $\tau$, because the contribution from the ionized circumgalactic/extragalactic medium already requires reionization be completed at an interestingly modest redshift. Ideas about ionizing radiation from sources present during the dark ages---maybe decaying dark matter, maybe primeval black holes that are evaporating or accreting baryons---are constrained by the condition that they not  make an unacceptable contribution to $\tau$ \cite{21cm_absorption}. That is, $\tau$ is a useful constraint on the search for ideas about more interesting dark ages \cite{DDm&tau}.

\item The CMB carries a rich memory of what happened during the course of cosmic evolution. That includes the effects on the intensity spectrum of the recombination radiation from decoupling and from the interaction of the CMB with hotter and colder free electrons in galaxies, clusters of galaxies, and the circumgalactic/extragalactic medium. Detection of these effects would be valuable constraint on ideas about cosmic evolution, but detection depends on an improved understanding of the foreground intensity spectrum. The ARCADE~2  sky brightness temperatures \cite{ARCADE} at frequencies less than about 10 GHz are greater than expected from deep radio source counts \cite{MeerKAT}, an open puzzle, possibly to be resolved by identification of energetic developments in the intergalactic medium at reionization, maybe driven by a primeval magnetic field \cite{primevalB}. 

\item Reionization that is presumed to be caused by young star-forming galaxies is expected to be patchy, and the effect of the variation of the optical depth $\tau$ across the sky adds to the secondary CMB anisotropy from sources along the lines of sight to high redshift. Detection of the effect of patchy reionization would be an informative guide to how the universe became ionized after the dark ages, perhaps possible to distinguish from the effects of microwave sources and scattering by electrons that are hot or moving, the Sunyaev-Zel'dovich effects \cite{patchy_reion, kSZ}, because of the different frequency dependences.

\item Thermally produced neutrinos accompanying the CMB surely are present too; their actual detection will be an important check \cite{PTOLEMY}. In the empiricist philosophy of this essay the measurement must be attempted; something may turn up.

\end {enumerate}

\bigskip\noindent(\stepcounter{topic}\thetopic){\it The Very Early Universe}\medskip

\noindent A useful measure of a span of cosmic evolution is the count of e-foldings of redshift, defined as $\ln (1+z_1)/(1+z_2)$ for the separation of events at redshifts $z_1$ and $z_2$. Matter and the cosmological constant made equal contributions to the rate of expansion of the universe at $z=0.33$, or 0.28 e-foldings ago. Cosmic matter was reionized at about 2.5 e-foldings ago. The ``dark ages'' lasted some 4.5 e-foldings from reionization back to decoupling of baryonic matter and radiation, totaling 7 e-foldings back in time from now. The light isotopes were formed 21 e-foldings ago. Inflation would have ended, or a bounce of some sort occurred, some 60 or 70 e-folds back in time. If inflation is a good approximation, then that phase lasted at least 60 e-folds. 

These spans of cosmic evolution include relatively brief logarithmic intervals when events left fossil records: light isotope formation, decoupling, galaxy formation, formation of the solar system, origin of life on Earth. In the standard cosmology nothing new was happening during long spans of cosmic evolution measured in e-foldings. This is in part because we happen to have observations of fossil evidence of what was happening in particular intervals of cosmic evolution. There cannot have been so many effects yet to be detected to have disturbed the present degree of agreement of theory and the cosmological tests, but we know of some things that seem sure to have happened at e-foldings to be discovered, and there are interesting ideas about others. 

\begin{enumerate}[a.] 

\item The cosmological inflation picture of the early universe is not yet a fixed theory; it offers suggestions rather than predictions and the suggestions can be adjusted. But the suggestions people arrived at early on, that space sections are flat with close to Gaussian adiabatic departures from homogeneity with the power spectrum slightly tilted from scale-invariance, are impressively successful. This was seen to follow from the general idea of inflation in a natural way before these initial conditions were observationally established. In the empiricist philosophy of this essay this makes the case for inflation particularly interesting, though of course not yet persuasive. 

What might have happened before the inflation epoch that produced our universe is an open question. In particular, eternal inflation cannot trace arbitrarily far back into the past, for if so we would be seeing other universes in the multiverse. That is, inflation is incomplete. I don't count this as an argument against inflation, eternal or otherwise. Bear in mind that all our physics is incomplete. 

It is important that there are alternatives to inflation to complete the picture of cosmic evolution, notably ideas about cyclic and bouncing universes \cite{Ashtekar, Brandenberger,  Penrose, ekpyrotic}. They too are incomplete; like inflation they introduce postulates that have no provenance other than what is needed to frame the picture. Again, this is the way it has to be, a situation to be sorted out. The gravitational waves that would be expected if the rate of expansion of the universe during inflation were large enough might yet be detected. That would add considerable weight to the empirical case for inflation, but a continued lack of detection might only signify an upper bound on the characteristic energy associated with the rate of expansion during inflation. Some other empirical constraint on the physics of the very early universe may turn up. If not we can expect the community to settle on a persuasive nonempirical construction \cite{Dawid}.
 
\item Something produced the residual baryons, the dark matter, and the cosmological constant. Maybe something in the dark ages or earlier made the seeds for cosmic magnetic fields \cite{Ratra, PenTurok}, or black holes with interesting masses, perhaps in rare departures from Gaussian curvature fluctuations, possibly occasioned by first-order phase transitions \cite{primevalBHs}. Maybe something made trace isocurvature fluctuations in the primeval matter distribution \cite{preBBNS}. 

Maybe early generations of cosmic strings or domain walls ran across parts of space when they were causally connected. Recall the curious alignment of radio galaxies, and clusters of galaxies, with the extended plane of the Local Supercluster shown in Figure~\ref{Fig:histograms}, and the curious contrast to the absence of alignment of the LIRGs, galaxies exceptionally luminous in the infrared. Might something have imprinted seeds that encouraged formation of AGNs and clusters of galaxies but not LIRGs? Might an inhomogeneous baryon mass fraction account for the curiously empty Local Void?  These are vague thoughts, but the empiricist philosophy calls for obsessive attention to the search for and possible interpretation of phenomena that seem strange, not likely to be accidental, and maybe have something of interest to teach us. 

\item Serendipity helps. FRBs were overlooked for a long time; they have given us a measure of the baryon density in the circumgalactic/extragalactic medium.  Detectors for the search for proton decay gave us neutrinos from the supernova  1987a. LIGO gave us a source of the r-process elements, and black hole binaries with unexpected masses. Following Mr. Micawber's philosophy, we must expect more fossils from the early universe will turn up.

\end{enumerate}

\bigskip\noindent(\stepcounter{topic}\thetopic){\it Living Matter}\medskip

\noindent This is not a neglected aspect of cosmology, but it deserves an entry in this list simply to assert its place in cosmology. 

Might explorations of the planets and satellites in the solar system reveal traces of what we could be willing to term life that could not have come from Earth? We are informed that there likely are billions of planets around stars in our galaxy. We can be sure that all sorts of marvelous things are happening on the surfaces of all these planets, and we hunger for the slightest glimpses, as  from SETI detections. Something will turn up.

\bigskip\noindent(\refstepcounter{topic}\thetopic\label{AnthropicPrinciple})
{\it The anthropic principle}\medskip

\noindent Anthropic explanations have the flavor of just so stories, such as how the leopard got its spots. But how else can we account for the allowed contribution of zero-point energies to the cosmic vacuum energy density represented by the cosmological constant? We had to have flourished in a galaxy capable of containing the debris that was recycled through a few generations of stars to produce the heavy elements we require. But our universe of vast numbers of galaxies containing immense numbers of planets seems to be an excessive response to the anthropic condition. Inflation need not have produced the enormous numbers of galaxies seen in our universe to make the single one we need; smaller primeval curvature fluctuations would produce quite a few galaxies. Larger primeval fluctuations would produce mass concentrations so dense that passing stars would disrupt something like the solar system, but I imagine it would leave in the field the occasional relatively small galaxy that we need. 

Another way to think of this is that the universe is a tranquil place, by and large. There are spectacularly violent events in the centers of galaxies, and we now know about the violence of merging black holes, but such things are far more rare than needed for our comfort. Most of space contains matter moving with peculiar velocities $\lap 1000$~km~s$^{-1}$, meaning spacetime curvature fluctuations are quite small, $(v/c)^2\lap 10^{-5}$. Did we need such tranquility? Again, it seems to be an excessive response to the anthropic explanation of our existence.

The weak form of the anthropic principle is the assumption that reality operates in what we would consider a reasonable and logical way; no miracles. Dicke \cite{Dicke1961} introduced this weak form with his argument that the expanding universe has to be at least several billion years old to have allowed time for the formation of the chemical elements, then the formation of the solar system, the origin of life on earth, and the evolution of the species to one that takes an interest in such things. In the $\Lambda$CDM theory the expanding universe is older than the greatest stellar evolution ages; the weak principle applied to this cosmology passes the test so far. 

We have not been issued a guarantee that reality respects our notions of logic. That is, the weak anthropic principle is an assumption that could be falsified, which would be interesting. 

I take the strong form of the anthropic principle to be the postulate of a statistical ensemble of universes, a multiverse. We live in a universe from the ensemble with properties that allow our existence. This offers a direct way to account for the quantum zero-point energy density, which is good. It offers an explanation of why the baryons and antibaryons thermally produced in the early universe were not completely annihilated; we would not be in a universe where that happened. But is the strong anthropic principle needed to account for the local baryon excess? There are inventive thoughts about particle physics that would explain it. Might other inventive ideas account for the quantum vacuum energy density?

The strong anthropic principle offers predictions of a curious nature. It is to be applied when all else fails, ignored when its predictions are abundantly satisfied, and falsified only by a failure of the weak principle.

\bigskip\noindent(\stepcounter{topic}\thetopic){\it Cosmology and Quantum Physics} \medskip

\noindent The puzzle of the quantum vacuum energy density bears repeating in a broader context. Relativity gave us the speed of light, gravity gave us Newton's gravitational constant, and quantum physics gave us Planck's constant. Cosmology gave us Hubble's constant, which we can be sure is not a constant but is a key measure of the present state of our universe. These four parameters yield two characteristic times and two mass densities:
$$
H^{-1} \sim 10^{17}\hbox{ sec}, \quad 
{H^2\over G} \sim 10^{-28}\hbox{ g cm}^{-3},
$$
$$  
\sqrt{G\hbar\over c^5} \sim 10^{-43}\hbox{sec}, \quad 
{c^5\over G^2\hbar}\sim 10^{94}\hbox{ g cm}^{-3}.
$$
The quantities in the first line make qualitative sense; they are comparable to the ages of old stars and the cosmic mean mass density. That was a great comfort in the early days of physical cosmology. The second line brings to mind no ready relevance to the large-scale nature of the universe. Other quantities you can derive from $\hbar$ and the other parameters are relevant to properties of atoms and molecules, stars, the breaking of exact homogeneity that gave us galaxies, evaporating black holes, and spacetime foam. But so far the major contribution of $\hbar$ to empirical cosmology is the puzzle of the cosmological constant. Where shall we turn to? Must it be the anthropic principle?

\bigskip\noindent(\stepcounter{topic}\thetopic){\it The Theory of Everything}\medskip

\noindent Martin Rees \cite{Rees} rightly celebrates the concept of the multiverse as the next layer in the sequence of revolutions in understanding of the world around us. We can list many layers: the Ptolemy universe with the earth centered in the crystal spheres that hold the astronomical objects; the Copernican universe with the sun at the center; the Kapteyn universe centered on the Milky Way galaxy; Hubble's realm of the nebulae with no center; and the multiverse of which our universe is a speck. The layers of discovery go down in scale too. Henri Poincar\'e \cite{Poincare} remarked that the Mariotte/Boyle law is simple and wonderfully accurate for many gases, but these gas examined in sufficiently fine detail break up into the complex motions of enormous numbers of particles. Poincar\'e asked whether gravity examined in sufficiently fine detail also departs from the simplicity of Newton’s law into complex behavior. Poincar\'e suggested we consider that ``then again [there may be] the simple under the complex, and so on, without our being able to foresee what will be the last term.'' Underlying the particle physicists' ``theory of everything'' could be more layers of Poincar\'e's successive approximations. And why should the layers of structure on large scales end with multiverses? This grand hierarchical vision must end with nonempirical assessments of conjectures about the extremes of large and small scales that the world economy cannot afford to test. But there is immense room in between for continued empirical exploration.

\bigskip\noindent Acknowledgements

\medskip\noindent This essay grew out of a list of issues to be prepared for a lecture at the 2021 Canadian Association of Physicists Congress. Roya Mohayaee's suggestions and a thorough exchange of opinions with Subir Sarkar improved this essay. I am grateful for advice from Neta Bahcall and Michael Strauss on statistical measures of galaxy positions and motions, Marla Geha on dwarf galaxies, Justin Khoury and Neil Turok on gravity physics, Lyman Page on CMB phenomena, Scott Trager on galaxy bimodality, Eline Tolstoy on the stellar halo of the Milky Way, and many other colleagues for exchanges of thoughts through the years.

\end{document}